\documentclass{iau}
\usepackage{graphicx}

\newcommand{\los}{\emph{los}}

\title[A Bayesian Method for the Extinction] %% give here short title %%
{A Bayesian Method for the Extinction}

\author[Hai-jun Tian]   %% give here short author list %%
{Hai-Jun Tian$^{1,2}$, Chao Liu$^{2}$,
Jing-Yao Hu$^2$,
Yang Xu$^2$,
Xue-Lei Chen$^2$
}

\affiliation{$^1 $China Three Gorges University, Yichang, 443002. Email: {\tt hjtian@lamost.org}\\[\affilskip]
$^2$National Astronomical Observatories, Chinese Academy of Sciences, Beijing 100012}

\pubyear{2014}
\volume{306}  %% insert here IAU Symposium No.
\pagerange{119--126}
\jname{Statistical Challenges in 21st Century Cosmology}
\editors{Alan Heavens, Jean-Luc Starck \& Alberto Krone-Martins, eds.}
\begin{document}

\maketitle
\vspace{-1.3em}
\begin{abstract}
We propose a Bayesian method to measure the total Galactic extinction parameters, $R_V$ and $A_V$. Validation tests based on the simulated data indicate that the method can achieve the accuracy of around 0.01\,mag. We apply this method to the SDSS BHB stars in the northern Galactic cap and find that the derived extinctions are highly consistent with those from \cite{SFD98}. It suggests that the Bayesian method is promising for the extinction estimation, even the reddening values are close to the observational errors. 

\end{abstract}
\firstsection % if your document starts with a section,
\vspace{-1.35em}
\section{Introduction}
The Galactic interstellar extinction is attributed to the absorption and scattering of the interstellar medium, such as gas and dust grains\cite{draine03}. The extinction, as a function of the wavelength, is related to the size distribution and abundances of the grains. Therefore, it plays an important role in understanding the nature of the interstellar medium. In addition, the flux of extragalactic objects suffers from different extinction in different bands, which leads to some bias on the extragalactic studies (\cite{guy10}; \cite{tian11}). Hence, understanding the total interstellar extinction in each line of sight (hereafter \los) is crucial for accurate flux measurements.

The all-sky dust map can either be constrained by measuring interstellar extinction, or by employing a tracer of ISM, e.g., HI. One of the most broadly used dust maps was published by \cite{SFD98} (hereafter SFD), who derived it from the dust emission at 100\,$\mu$m and 240\,$\mu$m. Since then, many other works have claimed discrepancy with their results(\cite{arce99}; \cite{Dobashi05}; \cite{peek10}).

This paper propose an effective method to examine the extinction values of SFD using the BHB stars as tracer in the northern Galactic cap. BHB stars are luminous and far behind the dusty disk, which contribute to most of the interstellar extinction.

\vspace{-1.44em}
\section{Methods}\label{bayesian}
{\underline{\it Bayesian Method for Color Excess}}. The total Galactic extinction in a given \los\ is measured from the offset of the observed color indexes of the BHB stars from their intrinsic values.  A set of BHB stars,  which dereddened color indexes, $\{\bf{c_{k}}\}$ (where $k=1, 2,\dots,N_{BHB}$), are known, are selected as template stars. The reddening of a field BHB stars can then be estimated by comparing their observed colors with the templates. Given a \los\ $i$ with $N_i$ field BHB stars, the posterior probability of the reddening $\bf{E_i}$ is denoted as $p(\bf{E}_{i}|\{\bf{\hat{c}_{ij}}\},\{\bf{c_k}\})$, where ${\bf \hat{c}_{ij}}$ is the observed color index vector of the BHB star $j$ in the \los\ $i$, and $\bf{c_k}$ the intrinsic color index vector of the template BHB star $k$. According to the Bayes theorem, this probability can be written as
\begin{equation}\label{eq:bayes}
\footnotesize
p({\bf E_{i}}|\{\bf{\hat{c}_{ij}}\},\{\bf{c_k}\}) = p(\{\bf{\hat{c}_{ij}}\}|\bf{E}_{i}, \{\bf{c_k}\})P({\bf E}_{i}|\{\bf{c_k}\}).
\end{equation}

The right-hand side can now be rewritten as
\begin{equation}
\footnotesize
p({\bf E_{i}}|\{\bf{\hat{c}_{ij}}\},\{\bf{c_k}\})  =  \prod_{j=1}^{N_{i}} \sum_{k=1}^{N_{BHB}} p({\bf \hat{c}}_{ij}|{\bf E}_{i}, {\bf c}_{k})p({\bf E}_{i}).
\end{equation}

We assume that the likelihood $p({\bf \hat{c}_{ij}}|({\bf E_{i}}, \bf c_{k}))$ is a multivariate Gaussian  and so can be expressed as :
\begin{equation}
\footnotesize
p({\bf \hat{c}}_{ij}|{\bf E}_{i}, \bf c_{k}) =  \frac{1}{(2\pi|{\bf \Sigma}|)^{m/2}}\exp(-{\bf x}^{T}{\bf \Sigma}^{-1}{\bf x}),
\end{equation}
where ${\bf x} = {\bf E} + {\bf c}_{k} - \hat{{\bf c}}_{ij}$,  and ${\bf \Sigma}$ is the m$\times$m covariance matrix of the measurement of the color indexes of the star $j$,
\begin{equation}
\footnotesize
{\bf \Sigma}  =   \left[ \begin{array}{cccc}
\sigma_{u}^{2} + \sigma_{g}^{2} & -\sigma_{g}^{2}& 0 & 0 \\
-\sigma_{g}^{2} & \sigma_{g}^{2} + \sigma_{r}^{2} & -\sigma_{r}^{2} & 0 \\
0 & -\sigma_{r}^{2} & \sigma_{r}^{2} + \sigma_{i}^{2} & -\sigma_{i}^{2}\\
0 & 0 & -\sigma_{i}^{2} & \sigma_{i}^{2} + \sigma_{z}^{2}
\end{array} \right],
\end{equation} 
where the $\sigma_{u}$,$\sigma_{g}$, $\sigma_{r}$, $\sigma_{i}$, and $\sigma_{z}$ are the measurement uncertainties of the $u$, $g$, $r$, $i$, and $z$, respectively. 

{\underline{\it Least-squares Method for $R_V$  and $A_V$}}. After deriving the probability of the reddening in a \los, the most likely reddening values, 
\begin{equation}\footnotesize
E_i=(E(u-g),E(g-r),E(r-i),E(i-z)),
\end{equation} 
can be obtained from the probability density function (PDF). They can then be used to derive the $R_V$ and $A_V$ given an extinction model, such as \cite{CCM89} (hereafter CCM), from the following equations,
\begin{small}
\begin{eqnarray} \label{Extinc}
E(u-g) = ((a_{u} + \frac{b_{u}}{R_{V}}) - (a_{g} + \frac{b_{g}}{R_{V}}))*A_{V},\  \ \ \nonumber
E(g-r) = ((a_{g} + \frac{b_{g}}{R_{V}}) - (a_{r} + \frac{b_{r}}{R_{V}}))*A_{V}\ \ \nonumber\\
E(r-i) = ((a_{r} + \frac{b_{r}}{R_{V}}) - (a_{i} + \frac{b_{i}}{R_{V}}))*A_{V},\ \ \ 
E(i-z) = ((a_{i} + \frac{b_{i}}{R_{V}}) - (a_{z} + \frac{b_{z}}{R_{V}}))*A_{V}.\ \ \ 
\end{eqnarray}
\end{small}
These are linear equations for $A_V$ and $A_V/R_V$ and can be easily solved with a least-squares or $\chi^2$ method to find the best fit $A_V$ and $R_V$ for each BHB star. The averaged $A_V$ and $R_V$ in each \los\ are obtained from the median values of all the stars located in the \los; and the uncertainties can be estimated from the median absolute deviation. The terms $a_x$ and $b_x$ are given from CCM.
%%%%%%%%%%%%%%%%%%%%%%%%%%%%%%%%%%%%%%%%%%%%%%%%%%%%%%%%%%%%%%%%%%%%%%%%%%%%%%%%
\begin{figure}
\centering
\includegraphics[width=\textwidth, height=3cm, clip]{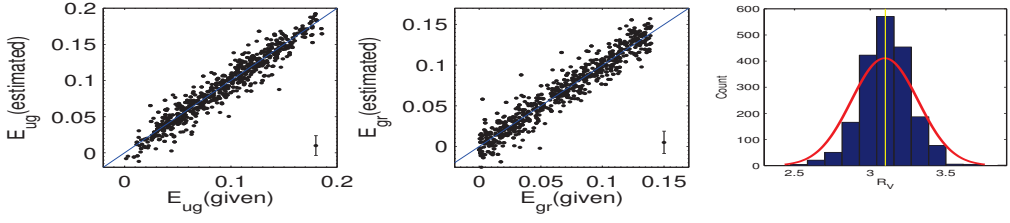}
\caption{The first two subplots show the comparison between the estimated and true values, the third presents the histogram of $R_V$.}
\label{mv}
\end{figure}
%%%%%%%%%%%%%%%%%%%%%%%%%%%%%%%%%%%%%%%%%%%%%%%%%%%%%%%%%%%%%%%%%%%%%%%%%%%%%%%%

{\underline{\it Validation of the Methods}}. We used 900 Monte Carlo simulations to validate the Bayesian method, the left and middle panels in Fig.\,\ref{mv} show the comparison between the estimated and the true extinction values (given in the simulations) in two colors. The mean 1$\sigma$ error bars (less than 0.01\,mag) are marked at the bottom, which suggests that the Bayesian method we employed in this work is robust.

To validate the least-squares method, we solve Eq.~\ref{Extinc} for the $E(B-V)$ data looked up from the SFD extinction maps for each \los\ and the fixed value of $R_V=3.1$. The right panel in Fig.\,\ref{mv} presents the histogram of $R_V$ in the simulation. The red curve is the Gaussian fit profile with the mean value $<R_V> \simeq 3.1$ and $\sigma \simeq 0.16$, the yellow line marks the location of $R_V = 3.1$.

\vspace{-1.4em}
\section{Application to the BHB stars}\label{bayesian}
\begin{figure}
\centering
\includegraphics[width=\textwidth, height=3cm, clip]{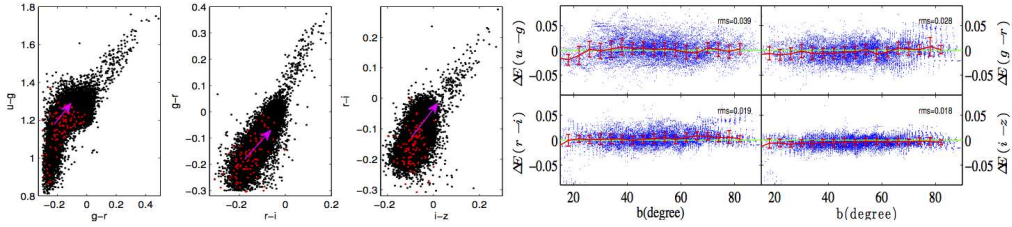}
\caption{Sample distribution in the 2-color space, and the reddening contrasts with SFD.}
\label{data_dist}
\end{figure}

{\underline{\it Data Selection}}. A total of 12 530 field BHB stars are selected from \cite{smith2010}, as shown with the black points in the left panel of Fig.\,\ref{data_dist}. The red points are the 94 zero-reddened template BHB stars selected from seven known globular clusters. The magenta arrows show the reddening direction.

{\underline{\it Reddening Values}}. The reddening values estimated by Bayesian method are compared with SFD in the right panel of Fig.\,\ref{data_dist}. They are well in agreement with each other.
%%%%%%%%%%%%%%%%%%%%%%%%%%%%%%%%%%%%%%%%%%%%%%%%%%%%%%%%%%%%%%%%%%%%%%%%%%%%%%%%
%%%%%%%%%%%%%%%%%%%%%%%%%%%%%%%%%%%%%%%%%%%%%%%%%%%%%%%%%%%%%%%%%%%%%%%%%%%%%%%%

\begin{figure}
\centering
\includegraphics[width=\textwidth, height=3cm, clip]{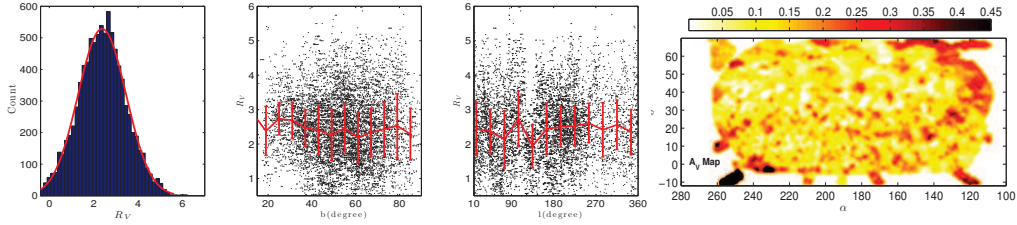}
\caption{Distributions of the measured $R_V$(the first three subplots), and $A_V$ map.}
\label{RvAv}
\end{figure}

{\underline{\it $R_V$ and $A_V$ Values}}. The left panel in Fig.\,\ref{RvAv} shows the histogram distribution of the measured $R_V$, best-fitted by a Gaussian with $\mu\simeq 2.4$ and $\sigma\simeq 1.05$. The middle two are the measured $R_V$ as a function of the Galactic latitude (the second panel) and longitude (the right panel), respectively. The red curves show $<R_V>$, which keeps constant at $\sim 2.5$ over all latitudes and longitudes. The right is the estimated $A_V$ map.
%%%%%%%%%%%%%%%%%%%%%%%%%%%%%%%%%%%%%%%%%%%%%%%%%%%%%%%%%%%%%%%%%%%%%%%%%%%%%%%%
%%%%%%%%%%%%%%%%%%%%%%%%%%%%%%%%%%%%%%%%%%%%%%%%%%%%%%%%%%%%%%%%%%%%%%%%%%%%%%%%

\vspace{-1.4em}
\section{Conclusions}
To measure the extinction, we propose a Bayesian method, and validated the method with simulations, which indicates accuracy is around 0.01\,mag. It is robust even in the case that the reddening values are close to the observational errors. The extinctions derived from the SDSS BHB stars with this method are high consistent with SFD.
%\section{Acknowledgments}

{\underline{\it Acknowledgments}}. The authors thank the grants (No. U1231123, U1331202, U1231119, 11073024, 11103027, U1331113, 11303020) from NSFC. THJ thanks the support from LAMOST Fellowship.

\end{document}